\documentclass[12pt]{article}
\newcommand{\secn}[1]{Section~\ref{#1}}
\newcommand{\bra}[1]{\langle{#1}|}
\newcommand{\ket}[1]{|{#1}\rangle}

\newcommand{\eq}[1]{Eq.~(\ref{#1})}

\def\beq{\begin{equation}}
\def\eeq{\end{equation}}
\def\beqa{\begin{eqnarray}}
\def\eeqa{\end{eqnarray}}
\newcommand{\sect}[1]{\setcounter{equation}{0}\section{#1}}
\renewcommand{\theequation}{\thesection.\arabic{equation}}
\newcommand{\EQ}{\begin{equation}}
\newcommand{\EN}{\end{equation}}
\newcommand{\bea}{\begin{eqnarray}}
\newcommand{\ena}{\end{eqnarray}}

\renewcommand{\a}{\alpha}

\newcommand{\NP}[1]{Nucl.\ Phys.\ {\bf #1}}
\newcommand{\PL}[1]{Phys.\ Lett.\ {\bf #1}}

\newcommand{\MPL}[1]{Mod.\ Phys.\ Lett.\ {\bf #1}}

\renewcommand{\thefootnote}{\fnsymbol{footnote}}
\def\one{{\hbox{ 1\kern-.8mm l}}}

\def\ii{{\rm i}}

\newlength{\bredde}
\def\slash#1{\settowidth{\bredde}{$#1$}\ifmmode\,\raisebox{.15ex}{/}
\hspace*{-\bredde} #1\else$\,\raisebox{.15ex}{/}\hspace*{-\bredde} #1$\fi}

\begin{document}
\begin{titlepage}
\rightline{DFTT 14/99}
\rightline{\hfill March 1999}
\vskip 1.2cm
\centerline{\Large \bf STABLE NON-BPS D-BRANES}
\vskip 0.5cm
\centerline{\Large \bf IN TYPE I STRING THEORY
\footnote{Work partially supported by the European Commission
TMR programme ERBFMRX-CT96-0045, by MURST and by the
Programme Emergence de la region Rh\^one-Alpes (France).
\\
E-mail:~~{\tt frau,gallot,lerda,strigazzi@to.infn.it}}}
\vskip 1.5cm
\centerline{\bf M. Frau $^a$, L. Gallot $^a$,
A. Lerda $^{b,a}$ and P. Strigazzi $^a$}
\vskip .8cm
\centerline{\sl $^a$ Dipartimento di 
Fisica Teorica, Universit\`a di Torino}
\centerline{\sl and I.N.F.N., Sezione di Torino, 
Via P. Giuria 1, I-10125 Torino, Italy}
\vskip .4cm
\centerline{\sl $^b$ Dipartimento di Scienze e Tecnologie Avanzate}
\centerline{\sl Universit\`a del Piemonte Orientale, I-15100 
Alessandria, Italy}
\vskip 1.5cm
\begin{abstract}
We use the boundary state formalism to study, from the
closed string point of view, superpositions of branes and anti-branes
which are relevant in some non-perturbative string
dualities. Treating the tachyon instability of these systems
as proposed by A. Sen, we show how to incorporate the effects of the
tachyon condensation directly in the boundary state. 
In this way we manage to show explicitly
that the D1 -- anti-D1 pair of Type I is a stable non-BPS D-particle,
and compute its mass. We also generalize this construction to describe
other non-BPS D-branes of Type I. By requiring the absence of tachyons
in the open string spectrum, we find which configurations are
stable and compute their tensions. Our classification is in
complete agreement with the results recently obtained using the K-theory
of space-time.
\end{abstract}
\end{titlepage}
\newpage
\renewcommand{\thefootnote}{\arabic{footnote}}
\setcounter{footnote}{0}
\setcounter{page}{1}
\sect{Introduction}
\label{intro}

Among the various string dualities, the one between
the SO(32) heterotic string and the Type I
theory \cite{WIT} is of special interest for many reasons.
First of all, this weak/strong coupling
duality relates two models with radically different
perturbative expansions and spectra.
Secondly, an analysis of the web of string dualities shows
that all of them may be ``derived'' by combining it with
T-duality \cite{sen0}.

The duality conjecture between the SO(32) heterotic string
and the Type I theory basically relies on the
uniqueness of the corresponding supergravity models.
In fact, the two tree-level effective actions turn out to
be related \cite{WIT}
through a field redefinition which inverts the
coupling constant. This duality thus identifies the weak
coupling regime of one theory with the strong coupling
regime of the other.
As a first check of the conjecture,
it has been shown \cite{DAB} that the
world-volume theory on the D-string of Type I is identical
to the world-sheet theory of the heterotic string. In particular
this explains the appearance of the SO(32) spinor
representation in the Type I theory.

More evidence of this conjectured duality has been 
given in the literature, mainly 
relying on the existence of BPS states. As is well known,
BPS states form short (or ultra-short) multiplets 
of the supersymmetry algebra 
for which the mass is related to the charge. 
Because of this fact they are stable and protected 
from radiative corrections. Hence, their properties 
can be analysed and computed perturbatively at 
weak coupling and then reliably extrapolated at 
strong coupling. Such tests of the heterotic/Type I duality conjecture 
comprise comparison between BPS spectra in the 
compactified theories \cite{GAVA}, but also 
between BPS saturated terms in the effective actions 
in both the uncompactified \cite{TSE} and the 
compactified cases \cite{BAC1}.

However, at the first massive level the heterotic string contains  
perturbative states which are stable but not BPS.
Their stability follows from the fact that they are the lightest states
carrying the quantum numbers of the spinor representation of SO(32).
Since they cannot decay, these states should be 
present also in the strong coupling regime. Then, if the 
heterotic/Type I duality is correct, the Type I theory should support
non-perturbative stable 
configurations that are spinors of SO(32).
Finding them and checking their multiplicities
is therefore a very non-trivial test on the heterotic/Type I duality.

It turns out \cite{sen4} that a pair formed by a 
D1-brane and an anti D1-brane of Type I 
(wrapped on a circle and with a 
$Z_2$ Wilson line) describes a configuration 
with the quantum numbers of the spinor representation of
SO(32) (for reviews see \cite{lr}). Thus this system is the right 
candidate to describe in the non-perturbative regime the stable 
non-BPS states of the heterotic string mentioned above.
However, a superposition of a brane with an anti-brane is unstable
due to the presence of tachyons in the open strings
stretching between the brane and the anti-brane \cite{BanksSuss}.
Then the problem of defining properly this superposition and
treating its tachyon instability arises.
This has been addressed by A. Sen in a remarkable series of 
papers \cite{sen1,sen3,sen4,sen5,sen6}. In particular in 
Ref.~\cite{sen4} he considered a D-string -- anti D-string
pair of Type IIB, and managed to prove that when the tachyon 
condenses to a kink along the compact direction of the
D-string, the pair becomes tightly bound and, as a whole,
behaves as a D-particle. He also computed its mass
finding that it is a factor of $\sqrt{2}$ bigger than the one of the 
supersymmetric BPS D-particle of the Type IIA theory. 
The presence of a D-particle in the Type IIB spectrum looks
surprising at first sight since one usually thinks that in Type IIB
there are only D$p$-branes with $p$ odd. However, one should keep in mind that
such a D-particle is a non-supersymmetric and non-BPS configuration.
Furthermore it is unstable, due to the fact that there are
tachyons in the spectrum of the open strings living on
its world-line. These tachyons turn out to be odd under the
world-sheet parity $\Omega$, and hence disappear if one performs
the $\Omega$ projection to get the Type I string \cite{ps}.
Therefore, the D-particle found by Sen is a stable non-perturbative 
configuration of Type I that transforms as a spinor of SO(32).

In this paper we analyze the Type I D-particle from the
point of view of the closed string operator formalism, and
construct its representation in terms of a boundary state
\cite{boundary,PCAI}.
The boundary state approach is a very convenient and 
useful method to describe BPS D-branes,
find their couplings with the various closed string fields
\cite{cpb} and also determine their interactions (for some applications
of the boundary state see for example Refs.~\cite{CANGEMI,BILLO,06}). 
In Refs.~\cite{sen3,Gab1}, it was shown 
that the boundary state formalism can also be applied successfully
to describe non-BPS configurations in certain orbifold models.

Here, we use this approach to study
the superposition of a D-string and an anti D-string
in the Type IIB theory, and construct the corresponding
boundary state with the proper normalization. 
Then, extending Sen's arguments, we manage
to incorporate the effects of the tachyon condensation directly
in the boundary state, and show explicitly that at a particular value 
of the tachyon v.e.v. this boundary state describes a non-BPS D-particle. 
Moreover, we show that the same result can be achieved by means
of a suitable discrete transformation which effectively accounts for the 
tachyon condensation on the initial D-string -- anti D-string pair.
Using this construction we can also compute directly the mass
of the non-BPS D-particle finding agreement with Sen's result.  

In the closed string operator formalism, the $\Omega$
projection that reduces Type IIB to Type I is implemented
by adding to the boundary state of Type II the 
crosscap state \cite{crosscap,PCAI} which
is uniquely determined by the requirement
of tadpole cancellation. In this way one obtains an
effective boundary state for the Type I D-branes
which can be efficiently used to study their properties
and interactions. Applying this procedure to the non-BPS
D-particle, we construct its corresponding boundary state of Type I
and use it to check its stability by showing that no
tachyon poles develop in the open string vacuum
amplitude at one loop.  

This construction can be easily generalized to
describe other non-BPS D-branes of Type I.
In particular, by requiring the absence of tachyons 
in the open string spectrum, we determine which of these
Type I configurations are stable and compute their tensions.
Our classification is in complete agreement 
with the results recently obtained by E. Witten \cite{WITTEN} and 
others \cite{HORAVAK,GUKK} using the K-theory of space-time.
Other recent papers discussing related subjects are \cite{Gab2,Yi,BGH}.

This paper is organized as follows: in \secn{boundarystate} we briefly
review the
boundary state formalism and specify our notations;
in \secn{d1antid1}, we derive the boundary state for the Type IIB D-particle
as a bound state of D-strings.
In \secn{typeI} we  
perform the $\Omega$ projection on the boundary state
to describe the D-particle of the Type I theory,
and explicitly verify its stability.
In \secn{DpI} we generalize these results to construct the boundary
state for non-BPS D$p$-branes of Type I and discuss for which values
of $p$ the tachyon cancellation
condition can be satisfied.
Finally, a few more technical calculations are reported in two 
appendices.

\vskip 1.5cm
\sect{The boundary state formalism}
\label{boundarystate}
In the closed string operator formalism
one describes the supersymmetric D$p$-branes of Type II 
by means of boundary states $\ket{Dp}$ \cite{boundary,PCAI}. 
These are 
closed string states which insert a boundary on the world-sheet, enforce
on it the appropriate boundary conditions and represent the 
source for the closed strings emitted by the branes 
\cite{cpb}.
In the fermionic string, both in the NS-NS 
and in the R-R sectors, there are two possible 
implementations 
for the boundary conditions of a D$p$-brane which
correspond to two boundary states $\ket{Dp, \pm}$. 
However, only the combinations
\begin{equation}
\ket{Dp}_{\rm NS} = \frac{1}{2} \Big[ \ket{Dp,+}_{\rm NS} 
- \ket{Dp,-}_{\rm NS} \Big]
\label{gsons}
\end{equation}
and
\begin{equation}
\ket{Dp}_{\rm R} = \frac{1}{2} \Big[ \ket{Dp,+}_{\rm R} + 
\ket{Dp,-}_{\rm R} \Big]
\label{gsor}
\end{equation}
are selected by the GSO projection in the NS-NS and in the R-R sectors
respectively. As discussed in Ref.~\cite{BILLO}, the boundary 
states $\ket{Dp,\pm}$ can be written as the product of a matter part 
and a ghost part
\begin{equation}
\label{bs001}
\ket{Dp, \pm } = \frac{T_p}{2}\,\ket{Dp_{\rm mat}, \pm } \ket{Dp_{\rm g}, 
\pm}~~, 
\end{equation}
where
\begin{equation}
\ket{Dp_{\rm mat}, \pm} = \ket{Dp_X} 
\ket{Dp_{\psi}, \pm}~~~,~~~
\ket{Dp_{\rm g}, \pm} = \ket{Dp_{\rm gh}} \ket{Dp_{\rm sgh}, \pm}~~.
\label{bs000}
\end{equation}
The overall normalization $T_p$ can be unambiguously fixed from
the factorization of amplitudes of closed strings emitted
from a disk \cite{FRAU,cpb} and is related to the brane tension $\tau_p$ 
\cite{lectPOL} according to
\begin{equation}
T_p= 
\tau_p\,\kappa_{10} =\sqrt{\pi}\left(2\pi\sqrt{\alpha'}\right)^{3-p}
\label{tension}
\end{equation}
where $\kappa_{10}$ is the gravitational coupling
constant in ten dimensions.
The explicit expressions of the various components of $\ket{Dp,\pm}$, 
together with their derivation, can be found for example
in Ref.~\cite{BILLO}. Here
we just recall the structure of the matter part of the boundary state 
\footnote{The ghost
and superghost parts will not play any significant role 
in our present analysis
and thus we omit them to avoid clutter.}, 
namely 
\begin{equation} \label{bs100}
\ket{Dp_X} = \exp\biggl[-\sum_{n=1}^\infty \frac{1}{n}\,
\a_{-n}\cdot S^{(p)}\cdot
\tilde \a_{-n} \biggr]\,\ket{Dp_X}^{(0)}~~,
\end{equation}
and
\begin{equation}
\label{bs101}
\ket{Dp_\psi,\pm}_{\rm NS} =  \exp\biggl[\pm\,\ii\sum_{r=1/2}^\infty
\psi_{-r}\cdot S^{(p)}\cdot\tilde \psi_{-r}\biggr]
\,\ket{0}~~,
\end{equation}
for the NS-NS sector, and
\begin{equation}
\label{bs102}
\ket{Dp_\psi,\pm}_{\rm R} = \exp\biggl[\pm\,\ii\sum_{n=1}^\infty
\psi_{-n} \cdot S^{(p)}\cdot\tilde \psi_{-n}\biggr]
\,\ket{Dp,\pm}_{\rm R}^{(0)}~~,
\end{equation}
for the R-R sector.
The matrix $S^{(p)}$
encodes the boundary conditions which characterize 
the D-brane; in the
case of a static brane without any external fields 
one simply has
\begin{equation}
S^{(p)}_{\mu\nu} = (\eta_{\alpha \beta}, -\delta_{ij})
\label{smatrix}
\end{equation}
where the indices $\alpha$, $\beta$ label 
the $(p+1)$ longitudinal directions
of the world-volume, and the indices $i$, $j$ 
label the $(9-p)$
transverse directions. Finally, the superscript $^{(0)}$ 
in Eqs. (\ref{bs100}) and (\ref{bs102}) denotes
the zero-mode contributions to the boundary state. 
These are
\begin{equation}
\ket{Dp_X}^{(0)} = \delta^{(9-p)}(q^i-y^i)\,\prod_{\mu=0}^9\ket{k^\mu=0}
\label{x0}
\end{equation}
where $y^i$ are the coordinates of the D-brane 
and $\ket{k^\mu}$ is the Fock vacuum
of momentum $k^\mu$, and 
\begin{equation}
\ket{Dp,\pm}_{\rm R}^{(0)} = \left(C\Gamma^0\ldots\Gamma^p
\frac{1\pm \ii\Gamma_{11}}{1\pm\ii}\right)_{AB} \ket{A} \ket{{\tilde B}}
\label{psi0}
\end{equation}
where $\ket{A}$ and $\ket{{\tilde B}}$ are the left 
and right spinor vacua
of the R sector
in the 32-dimensional Majorana representation 
\footnote{We refer to Refs.
\cite{BILLO,cpb} for the conventions on spinors 
and $\Gamma$ matrices.}. 
\par
Both in the NS-NS and in the R-R sectors, the two 
boundary states $\ket{Dp,\pm}$ are mapped into each 
other by the world-sheet fermion number operators $(-1)^F$
and $(-1)^{\widetilde F}$. In particular, in the 
NS-NS sector one can show that
\begin{equation}
(-1)^F\,\ket{Dp,+}_{\rm NS} = (-1)^{\widetilde F}\,
\ket{Dp,+}_{\rm NS}
= - \ket{Dp,-}_{\rm NS}
\label{dp+dp-}
\end{equation}
where, as usual, we have taken the NS-NS vacuum to 
be odd under $(-1)^F$ and $(-1)^{\widetilde F}$. 
Similarly, in the R-R sector, where the fermion
number operators also measure the chirality of the 
spinor vacua, one has \cite{BILLO,cpb}
\begin{equation}
(-1)^F\,\ket{Dp,+}_{\rm R} = (-1)^p\, \ket{Dp,-}_{\rm R} 
~~~~,~~~~
(-1)^{\widetilde F}\,\ket{Dp,+}_{\rm R}
= \ket{Dp,-}_{\rm R}~~.
\label{dp+dp-1}
\end{equation}
These relations become useful when one performs the
GSO projection. For example, in the NS-NS sector where
the GSO projector is
\begin{equation}
P_{\rm GSO} = \frac{1+(-1)^F}{2}\,
\frac{1+(-1)^{\widetilde F}}{2}~~,
\label{pgso}
\end{equation}
one simply has
\begin{equation}
\ket{Dp}_{\rm NS} \equiv P_{\rm GSO}\,\ket{Dp,+}_{\rm NS} 
= \frac{1}{2} \Big[ \ket{Dp,+}_{\rm NS} 
- \ket{Dp,-}_{\rm NS} \Big]~~,
\label{newdpns}
\end{equation}
{\it i.e.} \eq{gsons}. For the R-R sector, one has
to remember that the boundary states are written in the
left-right asymmetric superghost picture $(-1/2,-3/2)$
(see Ref.~\cite{BILLO} for details) in which the appropriate
GSO projector takes the form
\begin{equation}
P_{\rm GSO} = \frac{1+(-1)^p\,(-1)^F}{2}\,
\frac{1+(-1)^{\widetilde F}}{2}
\label{pgso1}
\end{equation}
where $p$ is even for Type IIA and odd for Type IIB in 
accordance with the R-R charge carried by a D$p$-brane.
Then, using \eq{dp+dp-1}, we easily see that
the GSO projected boundary state in the R-R sector
is
\begin{equation}
\ket{Dp}_{\rm R} \equiv P_{\rm GSO}\,\ket{Dp,+}_{\rm R} 
= \frac{1}{2} \Big[ \ket{Dp,+}_{\rm R} 
+ \ket{Dp,-}_{\rm R} \Big]~~,
\label{newdpr}
\end{equation}
{\it i.e.} \eq{gsor}.
This analysis shows that it is enough to consider the boundary
states $\ket{Dp,+}_{\rm NS}$ and $\ket{Dp,+}_{\rm R}$
from which the complete boundary states are obtained by means
of the GSO projection.
Finally, we recall that an anti D$p$-brane is a brane
with negative R-R charge, and in our formalism it is
simply described by a boundary state with an overall minus sign 
in the R-R sector.

Let us now consider the case in which the $p$ 
longitudinal space directions
are compactified on circles of radii $R$. The boundary 
state for a wrapped
D$p$-brane is still given by the previous expressions 
with only two
changes. The first (and easiest) one is in 
$\ket{Dp_X}^{(0)}$ which becomes
\begin{equation}
\ket{Dp_X}^{(0)} = \delta^{(9-p)}(q^i-y^i)\,\ket{k^0=0}
\prod_{\alpha=1}^p\sum_{w^\alpha}\ket{n^\alpha=0,w^\alpha}
\prod_{i=p+1}^9
\ket{k^i=0}
\label{x0comp}
\end{equation}
where $\ket{n^\alpha,w^\alpha}$ is the bosonic vacuum
with Kaluza-Klein index $n^\alpha$ and winding number 
$w^\alpha$.
T-duality invariance requires that these vacuum 
states must
be normalized as 
\begin{equation}
\bra{n',w'}n,w\rangle = \Phi\,\delta_{nn'}\delta_{ww'}
\label{norm}
\end{equation} 
where $\Phi$ is the self-dual ``volume'' of a compact 
direction~\footnote{The explicit expression of $\Phi$ as well as a
detailed discussion of its properties can be found 
for example in Ref.~\cite{GRV}.} which satisfies the following 
properties
\begin{equation}
\Phi \sim 2\pi R ~~{\mbox{for}}~R\to\infty~~~,~~~
\Phi \sim\frac{2\pi\alpha'}{R}~~{\mbox{for}}~R\to 0~~.
\label{volphi}
\end{equation}
As a consequence of this fact, a
second (and less obvious) change occurs in the boundary 
state of a wrapped D-brane. Indeed, as shown 
in Ref.~\cite{FRAU}, 
the overall normalization factor 
$T_p$ in \eq{bs001} must be replaced by  
\begin{equation} 
T_p \,\left(\frac{2\pi R}{\Phi}\right)^{p/2}~~.
\label{newtp}
\end{equation}
Then, it is easy to check that the boundary state 
defined in this way
correctly reproduces the vacuum amplitude for a 
wrapped D$p$-brane, normalization
factors included.

We conclude this brief summary by mentioning that in 
the boundary state formalism the
incorporation of $U(1)$ Wilson lines for wrapped 
D-branes amounts 
simply to introduce
in \eq{x0comp} appropriate phases
in the sum over all winding numbers of the compact 
directions, so that
$\ket{Dp_X}^{(0)}$ becomes
\begin{equation}
\ket{Dp_X}^{(0)} = \delta^{(9-p)}(q^i-y^i)\,\ket{k^0=0}
\prod_{\alpha=1}^p\sum_{w^\alpha}{\rm e}^{\ii\,\pi\theta_\a 
w^\a}\ket{n^\alpha=0,w^\alpha}\prod_{i=p+1}^9 \ket{k^i=0}
\label{x0wil}
\end{equation}
where the constants $\theta_\a$ parametrize the 
Wilson lines associated with the
$U(1)$ gauge fields living on the D-brane world-volume \cite{GREEN,FRAU}.

\vskip 1.5cm
\sect{The Type IIB D-particle as a bound state of D-strings}
\label{d1antid1}
Following Ref.~\cite{sen4}, we now study the Type IIB
system formed by the superposition of a D-string
and an anti D-string, both wrapped on a circle 
of radius $R$ 
and with a ${\bf Z_2}$
Wilson line on one of them, say for definiteness on the anti D-string. 
To describe this system we introduce the following 
boundary states
\begin{eqnarray}
\ket{B, +}_{\rm NS} &\equiv& \ket{D1, +}_{\rm NS} + 
\ket{D1', +}_{\rm NS}
\label{b1ns}\\
\ket{B, +}_{\rm R} &\equiv& \ket{D1, +}_{\rm R} - 
\ket{D1', +}_{\rm R}
\label{b1r}
\end{eqnarray}
where the $'$ indicates the presence of the ${\bf Z}_2$ 
Wilson line 
({\it i.e.} $\theta=1$ in the notation of \eq{x0wil}). 
Note that the 
minus sign in \eq{b1r} accounts for the fact that one of the
two members of the pair is an anti D-string.
Using the explicit expressions reported in Section 2, 
we have
\begin{eqnarray}
\ket{B,+}_{\rm NS} &=&
\frac{T_1}{2}\,\sqrt{\frac{2\pi R}{\Phi}}
\,\exp\biggl[-\!\sum_{n=1}^\infty \frac{1}{n}\,
\a_{-n}\cdot \hat{S}^{(1)}\cdot \tilde \a_{-n}\biggr]
\exp\biggl[ +\,\ii \!\sum_{r=1/2}^\infty
\psi_{-r}\cdot \hat{S}^{(1)}\cdot\tilde \psi_{-r}\biggr]
\nonumber \\
&&\exp\biggl[-\!\sum_{n=1}^\infty \frac{1}{n}\,
\a_{-n}\tilde \a_{-n}\biggr]
\exp\biggl[ +\,\ii\! \sum_{r=1/2}^\infty
\psi_{-r}\tilde\psi_{-r}\biggr] \ket{\Omega}_{\rm NS}
\label{d1ns}
\end{eqnarray}
where we have denoted by ${\hat S}^{(1)}$ the 
D-string $S$-matrix 
for all non-compact directions and have separately 
indicated in the 
second line
the contribution of the bosonic and fermionic 
non-zero modes of 
the compact direction ({\it i.e.} the modes of $X$, $\psi$ and
$\widetilde\psi$). Due to the presence of the ${\bf Z}_2$ 
Wilson line,
the vacuum $\ket{\Omega}_{\rm NS}$ is given by
\begin{eqnarray}
\ket{\Omega}_{\rm NS} &=& \delta^{(8)}(q^i)\ket{k^0=0}
\left(\sum_{w}\ket{0,w}
+\sum_{w}(-1)^w\ket{0,w}\right)\prod_{i=2}^9\ket{k^i=0}
\nonumber \\
&=&2 \,\delta^{(8)}(q^i)\ket{k^0=0}\sum_{w}\ket{0,2w}
\prod_{i=2}^9\ket{k^i=0}
\label{omegans}
\end{eqnarray}
where for simplicity we have set to zero the 
coordinates $y^i$ of the 
D-strings. Analogously, in the R-R sector we have
\begin{eqnarray}
\ket{B,+}_{\rm R} &=&
\frac{T_1}{2}\,\sqrt{\frac{2\pi R}{\Phi}}
\,\exp\biggl[-\!\sum_{n=1}^\infty \frac{1}{n}\,
\a_{-n}\cdot \hat{S}^{(1)}\cdot \tilde \a_{-n}\biggr]
\exp\biggl[ +\,\ii \sum_{n=1}^\infty
\psi_{-n}\cdot \hat{S}^{(1)}\cdot\tilde \psi_{-n}\biggr]
\nonumber \\
&&\exp\biggl[-\!\sum_{n=1}^\infty \frac{1}{n}\,
\a_{-n}\tilde \a_{-n}\biggr]
\exp\biggl[ +\,\ii \sum_{n=1}^\infty
\psi_{-n}\tilde\psi_{-n}\biggr] \ket{D1,+}^{(0)}_{\rm R}\,
\ket{\Omega}_{\rm R}
\label{d1r}
\end{eqnarray}
where $\ket{D1,+}^{(0)}_{\rm R}$ is defined 
in \eq{psi0} with $p=1$,
and 
\begin{eqnarray}
\ket{\Omega}_{\rm R} &=& \delta^{(8)}(q^i)\ket{k^0=0}
\left(\sum_{w}\ket{0,w}
-\sum_{w}(-1)^w\ket{0,w}\right)\prod_{i=2}^9\ket{k^i=0}
\nonumber \\
&=&2 \,\delta^{(8)}(q^i)\ket{k^0=0}\sum_{w}\ket{0,2w+1}
\prod_{i=2}^9\ket{k^i=0}~~.
\label{omegar}
\end{eqnarray}

Let us now suppose that the radius $R$ of 
the compact direction $X$ is given by 
\begin{equation}
R_c = \sqrt{\frac{\alpha'}{2}}~~.
\label{rcrit}
\end{equation}  
As shown in Ref.~\cite{sen4}, at this particular 
value of the radius
all excitations of the open strings stretched between 
the D-string
and the anti D-string have non negative mass squared. 
Moreover, at $R=R_c$
the bosonic coordinate $X$ is equivalent to a pair of 
fermionic fields $\xi$ and
$\eta$. Indeed, if we write 
\begin{equation}
X(z,\bar z) = \frac{1}{2}\left(X_{\rm L}(z) + X_{\rm R}
(\bar z)\right)
\label{xlxr}
\end{equation}
where
\begin{eqnarray}
X_{\rm L}(z) &=& q_{\rm L} -\ii (2\alpha ') p_{\rm L} 
\ln z +\ii
\sqrt{2\alpha'}
\sum_{n\not =0} \frac{\a_n}{n} z^{-n}
\nonumber \\
X_{\rm R}(\bar z) &=& q_{\rm R} -\ii (2\alpha ') p_{\rm R} 
\ln \bar z +\ii
 \sqrt{2\alpha'} \sum_{n\not =0} \frac{\tilde \a_n}{n} 
\bar z^{-n}
\label{expans}
\end{eqnarray}
with
\begin{equation}
p_{\rm L} =\frac{1}{2}\left( \frac{n}{R_c}+
\frac{w R_c}{\a '}\right)
~~,~~
p_{\rm R} =\frac{1}{2}\left( \frac{n}{R_c}-
\frac{w R_c}{\a '}\right)
\label{plpr}
\end{equation}
then, it is possible to prove that
\begin{eqnarray}
\exp\left[{\pm\frac{\ii }{\sqrt{2\a '}} X_{\rm L}(z)}\right] &\simeq&
\frac{1}{\sqrt{2}}
\Big(\eta(z)\pm \ii \xi(z)\Big)
\nonumber \\
\exp\left[{\pm\frac{\ii }{\sqrt{2\a '}} 
X_{\rm R}(\bar z)}\right] 
&\simeq& \frac{1}{\sqrt{2}}
\Big(\widetilde\eta(\bar z)
\pm \ii 
\widetilde\xi(\bar z)\Big)~~. \label{csieta}
\end{eqnarray}
By recombining the fermions $\eta$, $\xi$ and 
$\psi$ (and $\widetilde\eta$, $\widetilde\xi$ 
and $\widetilde\psi$) 
in a different manner,
we can obtain an equivalent representation of 
the same conformal field theory in terms of a
new compact bosonic field 
\begin{equation}
\phi(z,\bar z) = \frac{1}{2}\Big(\phi_{\rm L}(z)+
\phi_{\rm R}(\bar z)
\Big)
\label{phi}
\end{equation}
defined through
\begin{eqnarray}
\frac{1}{\sqrt{2}}
\left(\xi(z)\pm \ii \psi(z)\right)
&\simeq&\exp\left[{\pm\frac{\ii }{\sqrt{2\a '}} 
\phi_{\rm _L}(z)}\right] \nonumber \\
\frac{1}{\sqrt{2}}\left(\widetilde\xi(\bar z)\pm \ii 
\widetilde\psi(\bar z)\right)
&\simeq&\exp\left[{\pm\frac{\ii }{\sqrt{2\a '}} 
\phi_{\rm R}(\bar z)}\right]~~.
\label{philphir}
\end{eqnarray}
The field $\phi(z,\bar z)$ is a free boson of radius 
$R=R_c$ which has 
a mode expansion similar
to the one of the original coordinate $X(z,\bar z)$ 
({\it i.e.} Eqs. (\ref{expans}) and (\ref{plpr})) with
oscillators $\phi_n$ and $\tilde\phi_n$, 
and with Kaluza-Klein and 
winding numbers $n_\phi$ and $w_\phi$ respectively.

As emphasized in Ref.~\cite{sen4}, if one uses 
the new fields $\phi$ and 
$\eta$ instead of the original $X$ and $\psi$, 
it is possible 
to study explicitly the effects of a non vanishing 
v.e.v. for tachyon of the open string stretched 
beween the D-string and the anti D-string. 
In fact, in the 0 superghost
picture such a tachyon is described by the 
following vertex operator
\begin{equation}
V(z) = \frac{\ii}{\sqrt{2\a'}} \partial\phi(z) 
\otimes \sigma^1
\label{tachyon}
\end{equation}
where we have denoted by $\phi(z)$ the {\it open} 
string field 
corresponding to \eq{phi} and by $\sigma^1$ the 
Chan-Paton factor
appropriate for an open string stretched across 
the D-string -- anti D-string pair \cite{sen4}.
As a matter of fact, the vertex operator (\ref{tachyon}) does 
not represent a true tachyon since it creates 
a state which, at the 
critical radius $R=R_c$, is massless. However, 
since such a state becomes tachyonic in the 
decompactification limit $R\to\infty$, 
the field associated to \eq{tachyon} is, nevertheless, 
called tachyon. From the explicit expression of $V(z)$, 
we easily see that giving a non vanishing v.e.v. 
to the tachyon field is
equivalent to introducing a $U(1)$ Wilson line along $\phi$,
which we can parametrize as follows
\begin{equation}
{\cal W}(\theta) = \frac{1}{2}{\rm Tr}\left[
\exp\left(\frac{\theta}{2} \oint dz 
\frac{\ii}{\sqrt{2\alpha'}}\partial\phi  
\otimes \sigma^1
\right) \right]~~.
\label{wilson}
\end{equation}
Since in a pure closed string amplitude 
there are no other sources
of Chan-Paton factors, \eq{wilson} simplifies to
\begin{equation}
{\cal W}(\theta) = \cos\left(
\frac{\pi\theta w_\phi}{2}\right)
\label{wilson1}
\end{equation}
where $w_\phi$ is the total winding number 
of the closed
string state seen by the operator ${\cal W}(\theta)$.
 
We are now in the position of writing the boundary state
which describes the D-string -- anti D-string pair in the
presence of a non vanishing tachyon v.e.v. This is given by
Eqs. (\ref{d1ns}) and (\ref{d1r}) with the oscillators
$\a_n$, $\tilde\a_n$, $\psi_r$, $\tilde\psi_r$ of the compact
direction replaced by $\phi_n$, $\tilde\phi_n$, $\eta_r$ and
$\tilde\eta_r$, and with a vacuum that carries an explicit
dependence on the parameter $\theta$ according to \eq{wilson1}. 
In particular, in the NS-NS sector we have
\begin{eqnarray}
\ket{B(\theta),+}_{\rm NS} \!\! &=&\!\!
\frac{T_1}{2}\sqrt{\frac{2\pi R_c}{\Phi}}
\exp\biggl[-\!\sum_{n=1}^\infty \frac{1}{n}\,
\a_{-n}\cdot \hat{S}^{(1)}\cdot \tilde \a_{-n}\biggr]
\exp\biggl[ +\,\ii \!\sum_{r=1/2}^\infty
\psi_{-r}\cdot \hat{S}^{(1)}\cdot\tilde \psi_{-r}\biggr]
\nonumber \\
&&\!\exp\biggl[-\!\sum_{n=1}^\infty \frac{1}{n}\,
\phi_{-n}\tilde \phi_{-n}\biggr]
\exp\biggl[+\ii\! \sum_{r=1/2}^\infty
\eta_{-r}\tilde\eta_{-r}\biggr] 
\ket{\Omega(\theta)}_{\rm NS}
\label{hatbns}
\end{eqnarray}
where
\begin{equation}
\ket{ \Omega(\theta)}_{\rm NS} = 
2 \,\delta^{(8)}(q^i)\ket{k^0=0}
\sum_{w_\phi}\cos(\pi\theta 
w_\phi)\,\ket{0,2w_\phi}\prod_{i=2}^9\ket{k^i=0}~~.
\label{omegahatns}
\end{equation}
Analogously, in the R-R sector we have
\begin{eqnarray}
\ket{ B (\theta),+}_{\rm R} \!&=&\!
\frac{T_1}{2}\sqrt{\frac{2\pi R_c}{\Phi}}
\,\exp\biggl[-\!\sum_{n=1}^\infty \frac{1}{n}\,
\a_{-n}\cdot \hat{S}^{(1)}\cdot \tilde \a_{-n}\biggr]
\exp\biggl[ +\,\ii \!\sum_{n=1}^\infty
\psi_{-n}\cdot \hat{S}^{(1)}\cdot\tilde \psi_{-n}\biggr]
\nonumber \\
&&\exp\biggl[-\!\sum_{n=1}^\infty \frac{1}{n}\,
\phi_{-n}\tilde \phi_{-n}\biggr]
\exp\biggl[+\ii\! \sum_{n=1}^\infty
\eta_{-n}\tilde\eta_{-n}\biggr] \ket{D1,+}^{(0)}_{\rm R}\,
\ket{\Omega(\theta)}_{\rm R}
\label{hatbr}
\end{eqnarray}
where
\begin{equation}
\ket{\Omega(\theta)}_{\rm R} = 
2 \,\delta^{(8)}(q^i)\ket{k^0=0}\sum_{w_\phi}
\cos\left(\pi\theta(w_\phi+\frac{1}{2})\right)\,
\ket{0,2w_\phi+1}\prod_
{ i = 2 }^9\ket{k^i=0}~~. \label{omegahatr}
\end{equation}
Notice that at $\theta=0$ the boundary 
states (\ref{hatbns}) and 
(\ref{hatbr}) are completely equivalent to 
the original ones written in 
Eqs. (\ref{d1ns}) and (\ref{d1r}), as one can 
check for example
by computing some correlation functions or the 
vacuum amplitude. 
In this respect, it is
worth pointing out that in computing amplitudes
with $\ket{B(\theta),+}_{\rm NS}$
and $\ket{B(\theta),+}_{\rm R}$, 
one has to be careful in performing
correctly the GSO projection. In fact, this is 
{\it not} obtained by 
taking linear combinations as in Eqs. (\ref{gsons}) 
and (\ref{gsor}),
since the operators $(-1)^F$ and $(-1)^{\widetilde F}$ 
are {\it not} related
to the $\eta$ and $\tilde \eta$ fermion numbers. 
Instead, as is clear 
from \eq{philphir}, in the NS-NS sector one has
\begin{equation}
(-1)^{\widetilde F} ~~:~~ \widetilde\a_{n}^\mu \to 
\widetilde\a_{n}^\mu~~,~~ \widetilde\psi_r^\mu 
\to -\widetilde\psi_r^\mu
~~,~~\widetilde\phi_n \to 
-\widetilde\phi_n~~,~~\widetilde\eta_r\to
\widetilde\eta_r 
\label{-1f}
\end{equation}
and similarly for $(-1)^{F}$ on the left moving 
oscillators.
Using these rules, one can easily see, for example, 
that
\begin{eqnarray}
(-1)^{\widetilde F}\,\ket{B(\theta),+}_{\rm NS}
&=&-
T_1\,\sqrt{\frac{2\pi R_c}{\Phi}}
\,\exp\biggl[-\!\sum_{n=1}^\infty \frac{1}{n}\,
\a_{-n}\cdot \hat{S}^{(1)}\cdot \tilde \a_{-n}\biggr]
\nonumber \\
&&\exp\biggl[ -\,\ii \!\sum_{r=1/2}^\infty
\psi_{-r}\cdot \hat{S}^{(1)}\cdot\tilde \psi_{-r}\biggr]
\label{-1fb} \\
&&\exp\biggl[+\sum_{n=1}^\infty \frac{1}{n}\,
\phi_{-n}\tilde \phi_{-n}\biggr]
\exp\biggl[+\ii\!\sum_{r=1/2}^\infty
\eta_{-r}\tilde\eta_{-r}\biggr] \nonumber \\
&&\delta^{(8)}(q^i)\ket{k^0=0}
\sum_{w_\phi}\cos(\pi\theta 
w_\phi)\,\ket{w_\phi,0}\prod_{i=2}^9\ket{k^i=0}
~~.\nonumber 
\end{eqnarray}
Notice that the 
action of the operator $(-1)^{\widetilde F}$
on the bosonic field $\phi$ given
in \eq{-1f} looks like a
T-duality because it amounts to change the 
relative sign between its 
left and right moving oscillators. However, since the
compactification radius of $\phi$ is such that
\begin{equation} 
R_c = \frac{\a'}{2R_c}~~,
\label{rdual}
\end{equation}
this change in sign implies that a state with Kaluza-Klein 
index 
$n_\phi$ and winding number $w_\phi$ is transformed into a state
with Kaluza-Klein index $w_\phi/2$ (which is acceptable only 
when $w_\phi$ is even) and winding number $2n_\phi$, 
that is
\begin{equation}
(-1)^{\widetilde F}~~:~~ \ket{n_\phi,w_\phi} ~\rightarrow
\ket{\frac{w_\phi}{2}, 2n_\phi}~~.
\label{nwexchange}
\end{equation}
This peculiar behavior explains the structure of 
the vacuum
in \eq{-1fb}. Of course, similar considerations 
apply also for the boundary states in the R-R sector.
 
Let us now turn to the vacuum amplitude of the theory 
defined
on the world-volume of our D-string -- anti D-string pair. 
In the boundary state
formalism this amplitude is simply given by
\begin{equation}
{\cal A}(\theta) = \bra{B(\theta),+}\,P_{\rm GSO}\,
D\,\ket{B(\theta),+}
\label{ampl}
\end{equation}
where the GSO projection operator is
given in Eqs. (\ref{pgso}) and (\ref{pgso1}), 
and $D$ is the closed string propagator
\begin{equation}
D = \frac{\a'}{4\pi} \int \frac{d^2z}{|z|^2}
\,z^{L_0-a}\,{\bar z}^{\widetilde L_0-a}
\label{propa}
\end{equation}
with intercept $a_{\rm NS}=1/2$ in the NS-NS sector, and 
$a_{\rm R}=0$
in the R-R sector. Using the explicit expressions of the 
boundary states written above, and performing 
standard manipulations, one finds
\begin{equation}
{\cal A}_{\rm NS-NS}(\theta)
= \frac{VR_c}{2\pi^2\a '}
\int_{0}^\infty \!dt \left(\frac{\pi}{t}\right)^4 \!\left[\!
\left(\sum_{w_\phi}\cos^2(\pi\theta w_\phi) ~q^{w_\phi^2}\right)
\!\frac{f_3^8(q)}{f_1^8(q)}
-\sqrt{2}\,\frac{f_4^7(q)\,f_3(q)}{f_1^7(q)
\,f_2(q)}
\right]
\label{vacampl}
\end{equation}
and
\begin{equation}
{\cal A}_{\rm R-R}(\theta) = - \frac{VR_c}{2\pi^2\a '}
\int_{0}^\infty dt \left(\frac{\pi}{t}\right)^4 
\left[\sum_{w_\phi}\cos^2\left(\pi\theta (w_\phi +\frac{1}{2})\right)
~q^{\left(w_\phi+\frac{1}{2}\right)^2}\right]
\frac{f_2^8(q)}{f_1^8(q)}
\label{vacampl1}
\end{equation}
where $V$ is the (infinite) length of the time direction 
and 
\begin{eqnarray}
f_1(q) &=& q^{1 \over 12} \prod_{n=1}^{\infty} (1-q^{2n})
~~~,~~~f_2(q) = \sqrt{2} q^{1 \over 12}
 \prod_{n=1}^{\infty} (1+q^{2n})\nonumber\\
f_3(q) &=& q^{-{1 \over 24}} \prod_{n=1}^{\infty} (1+q^{2n-1})
~~~,~~~
f_4(q)= q^{-{1 \over 24}} \prod_{n=1}^{\infty} (1-q^{2n-1})
\end{eqnarray}
with $q={\rm e}^{-t}$.

It is interesting to observe that the contribution of the NS-NS$(-1)^F$ spin
structure  ({\it i.e.} the second term in \eq{vacampl}) 
does not depend 
on the tachyon v.e.v. $\theta$. This is a direct 
consequence of the
fact that this  spin structure arises from the 
overlap between
$\ket{B(\theta),+}_{\rm NS}$, 
whose vacuum contains states with only
winding numbers, and $(-1)^{\widetilde F}
\ket{B(\theta),+}_{\rm NS}$, whose vacuum instead 
contains states with only Kaluza-Klein numbers (see
Eqs. (\ref{omegahatns}) and (\ref{-1fb})). Therefore, in the
NS-NS$(-1)^F$ spin structure there is no contribution from the
bosonic zero modes of the compact direction $\phi$, and hence
no dependence on the tachyon v.e.v. $\theta$. 

If one performs
the modular transformation $t \to \pi/t$, 
the entire amplitude ${\cal A}(\theta)$
can be interpreted as the one-loop vacuum energy of the open strings
living in the world-volume of the D-string -- anti D-string pair.
After this modular transformation, one can explicitly 
check that
${\cal A}(\theta)$ in \eq{ampl} coincides with the  
annulus amplitude that follows from the rules 
described in Ref.~\cite{sen4} from the open
string point of view (see Appendix A for some details). 
In particular, one sees that the $\theta$-independent
NS-NS$(-1)^F$ spin structure of the closed string 
channel goes into the
R spin structure of the open string channel, that 
indeed has been shown
in Ref.~\cite{sen4} to be independent of the tachyon v.e.v.
 
At $\theta=1$ a remarkable simplification 
occurs: the R-R contribution to ${\cal A}$
vanishes, since, as is clear from \eq{omegahatr}, 
the R-R boundary state is null
at $\theta=1$. Thus, the entire vacuum amplitude 
becomes 
\begin{eqnarray}
{\cal A}(\theta=1) &=& \frac{VR_c}{2\pi^2\a '}
\int_{0}^\infty \!dt \left(\frac{\pi}{t}\right)^4 \!
\left[\left(\sum_{w_\phi} q^{w_\phi^2}\right)
\frac{f_3^8(q)}{f_1^8(q)}
-\sqrt{2}\,\frac{f_4^7(q)\,f_3(q)}{f_1^7(q)\,f_2(q)}
\right]
\nonumber \\
&=& \frac{V}{4\pi^2R_c}
\int_{0}^\infty \!dt \left(\frac{\pi}{t}\right)^4 
\left(\sum_{w_\phi} q^{w_\phi^2}\right)
\left[\frac{f_3^8(q)}{f_1^8(q)}
-\frac{f_4^8(q)}{f_1^8(q)}\right]
\label{ampl1}
\end{eqnarray}
where in obtaining the last line we have used \eq{rdual} 
to rewrite
the prefactor, and exploited the identities
\begin{equation}
f_2(q)\,f_3(q)\,f_4(q) = \sqrt{2}~~~~,~~~~f_1(q)\,
f_3^2(q)
=\sum_{n=-\infty}^{+\infty} q^{n^2}
\label{ident}
\end{equation}
to transform the integrand. Notice that with these 
manipulations we have
managed to reconstruct the typical combination of 
$f$-functions
that is produced by the {\it usual} GSO projection 
of the NS-NS sector.
Thus, we are lead to suspect that a simpler 
underlying structure may actually
exist at $\theta=1$, in accordance with the property, 
shown in Ref.~\cite{sen4}, 
that at this value of the tachyon v.e.v.
the D-string -- anti D-string pair is most tightly 
bound and,
as a whole, behaves like a D-particle. Indeed, 
our expectation
is confirmed by the fact that 
the amplitude ${\cal A}(\theta=1)$
can be factorized in terms of a new boundary 
state according to
\begin{equation}
{\cal A}(\theta=1) = \bra{\widetilde B,+}\,P_{GSO}\,
D\,\ket{\widetilde B,+}
\label{ampl2}
\end{equation}
where
\begin{eqnarray}
\ket{\widetilde B,\pm} &=&
\frac{T_1}{2}\,\sqrt{\frac{\pi\a'}{R_c\Phi}}
\,\exp\biggl[-\!\sum_{n=1}^\infty \frac{1}{n}\,
\a_{-n}\cdot \hat{S}^{(1)}\cdot \tilde \a_{-n}\biggr]
\exp\biggl[ \pm\,\ii \!\sum_{r=1/2}^\infty
\psi_{-r}\cdot \hat{S}^{(1)}\cdot\tilde \psi_{-r}\biggr]
\nonumber \\
&&\!\exp\biggl[+\sum_{n=1}^\infty \frac{1}{n}\,
\a_{-n}\tilde \a_{-n}\biggr]
\exp\biggl[ \mp\ii\! \sum_{r=1/2}^\infty
\psi_{-r}\tilde\psi_{-r}\biggr] \ket{\widetilde\Omega}
\label{newbound}
\end{eqnarray}
with
\begin{equation}
\ket{\widetilde \Omega}= 
2 \,\delta^{(8)}(q^i)\ket{k^0=0}
\sum_{w}\ket{w,0}\prod_{i=2}^9\ket{k^i=0}~~.
\label{newomega}
\end{equation}
Of course, the simple factorization of a vacuum amplitude
does not allow to uniquely fix the structure of 
a boundary state,
and thus one may think that there is some arbitrariness
in \eq{newbound}. However, a detailed analysis
of correlation functions shows that the new boundary state
$\ket{\widetilde B,+}$, which is written in 
terms of the 
original degrees of freedom for the compact 
direction ({\it i.e.}
$X$ and $\psi$), is equivalent to the boundary state
of \eq{hatbns} for $\theta=1$. To see this, one considers
closed string vertex operators written in terms of $X$
and $\psi$ and computes the corresponding correlation 
functions on a disk using the boundary state
$\ket{\widetilde B,+}$; then, transforming 
$(X,\psi)$ into $(\phi,\eta)$ through the bosonization 
formulas, one
checks that the same correlation functions are obtained
with the boundary state $\ket{B(\theta=1),+}$ given
in \eq{hatbns}. A few examples of such calculations
are explicitly described in Appendix B. 

Based on these results, we can conclude that in order 
to describe the D-string -- anti D-string pair at $R=R_c$
in terms of $X$ and $\psi$, we have to use the original 
boundary states of Eqs. (\ref{d1ns}) and (\ref{d1r})
if $\theta=0$, whereas we have to use the 
the new boundary state of \eq{newbound}
if $\theta=1$.  Of course, at this particular 
value of $\theta$ there is no R-R sector, as we have 
explicitly shown. It is interesting to observe that
$\ket{B,+}_{\rm NS}$ and $\ket{\widetilde B,+}$ can
be related to each other by means of a discrete transformation
${\cal T}$. Indeed,
as is clear from Eqs. (\ref{d1ns}) and (\ref{newbound}),
we may go from $\ket{B,+}_{\rm NS}$ to 
$\ket{\widetilde B,+}$ by changing the sign to the
right moving oscillators of the compact direction, and
consequently by changing the vacuum from $\ket{0,2w}$ to 
$\ket{w,0}$ since the radius of $X$ satisfies \eq{rdual}
(cf also \eq{nwexchange}).
Like the usual T-duality, also ${\cal T}$ transforms a
longitudinal direction into a 
transverse one, so that the new boundary state
$\ket{\widetilde B,+}$ more properly describes a
D0-brane with a compact transverse direction.
However, unlike T-duality,
${\cal T}$ switches off the R-R sector.
This fact suggests that, more than a symmetry of the theory,
this transformation ${\cal T}$ has to be regarded simply as
an effective way of implementing the change of
the tachyon v.e.v. from $\theta=0$ to $\theta=1$ on the 
original boundary states, which can be justified
by introducing the new fields $\phi$ and $\eta$
through the bosonization procedure as we have done.

In Ref.~\cite{sen4} it was shown that the 
decompactification limit $R\to \infty$ of the D-string --
anti D-string pair is meaningful only at 
$\theta=1$, where the
system is most tightly bound and no tadpoles develop.
In our formalism, this feature is revealed by the fact
the limit $R\to\infty$ is ill-defined on the original 
boundary states $\ket{B,+}_{\rm NS}$ and 
$\ket{B,+}_{\rm R}$ since their vacuum contains 
only a subset of all possible winding states, namely 
the states
with only even or odd winding numbers respectively in the 
NS-NS and R-R sectors. On the contrary, there are no
problems in taking the decompactification
limit on the new
boundary state $\ket{\widetilde B,+}$ which 
describes the
theory at $\theta=1$. In fact, when
$R\to\infty$ we can rewrite 
\eq{newomega} as follows
\begin{eqnarray}
\ket{\widetilde\Omega} &=& 2\,
\delta^{(8)}(q^i)\,\ket{k^0=0}~2\pi R\int
\frac{dk^1}{2\pi}\ket{k^1}\,\prod_{i=2}^9\ket{k^i=0}
\nonumber \\
&=&4\pi R\,\delta^{(9)}(q^i)\,
\prod_{\mu=0}^9\ket{k^\mu=0}
\label{newomega0}
\end{eqnarray}
which resembles \eq{x0} for $p=0$. Furthermore,
combining the factor
of $4\pi R$ from \eq{newomega0} with the prefactors of
$\ket{\widetilde B,+}$, we see that
the complete normalization of the boundary state 
becomes
\begin{equation}
\frac{T_1}{2}\,\sqrt{\frac{\pi\a'}{R\Phi}}~
4\pi R ~~~\longrightarrow ~~~
\frac{\sqrt{2}\,T_1\,(2\pi\sqrt{\a'})}{2}
~=~\frac{\sqrt{2}\,T_0}{2}
\label{newt0}
\end{equation}
where we have used the asymptotic behavior of 
$\Phi$ for $R\to\infty$
(see \eq{volphi}) and the explicit expression of the
tensions $T_p$ (see \eq{tension}). 
Thus, in the decompactification
limit our system is described by
\begin{eqnarray}
\ket{\widetilde B,+} &=&
\frac{\sqrt{2}\,T_0}{2}
\,\exp\biggl[-\!\sum_{n=1}^\infty \frac{1}{n}\,
\a_{-n}\cdot {S}^{(0)}\cdot \tilde \a_{-n}\biggr]
\exp\biggl[+\,\ii \!\sum_{r=1/2}^\infty
\psi_{-r}\cdot {S}^{(0)}\cdot\tilde \psi_{-r}\biggr]
\nonumber \\
&&\delta^{(9)}(q^i)\,\prod_{\mu=0}^9\ket{k^\mu=0}~~.
\label{bound0}
\end{eqnarray}
By performing the usual GSO projection, we then
obtain the complete boundary state
\begin{equation}
\ket{\widetilde B} \equiv P_{\rm GSO}\,
\ket{\widetilde B,+}= \frac{1}{2}\Big[\ket{\widetilde
B,+} -\ket{\widetilde B,-}\Big]
\label{btilde}
\end{equation}
which describes a D0-brane
in the Type IIB theory. Since there is no R-R sector, 
this D-particle is 
non-supersymmetric and non-BPS. Moreover, from \eq{bound0}
we explicitly see that
its tension (or mass) 
is a factor of $\sqrt{2}$ bigger than
the tension of the ordinary supersymmetric D-particle
of the Type IIA theory.
This fact has been 
proved and checked also in Ref.~\cite{sen4} 
using completely different arguments.

\vskip 1.5cm
\sect{The boundary state for the Type I D-particle}
\label{typeI}
The non-supersymmetric D-particle described in the
previous section is an unstable configuration
of the Type IIB theory. In fact, the absence of the
R-R part in the D-particle boundary state $\ket{\widetilde B}$
implies the absence of any
GSO projection in the dual open string theory defined
on its world-line. Thus, the spectrum of these
open strings contains not only the states that
a supersymmetric D-particle would sustain, but also
all other states that are usually removed by the GSO
projection. In particular in the NS sector, one finds
a tachyonic state which is responsible for the
instability of the entire configuration.
However, things might improve if there is a
consistent truncation of the theory which is
free of tachyons. One possibility is to study
our system in the Type I theory, which 
can be viewed as an orbifold of the 
Type IIB theory with respect to 
the world-sheet parity $\Omega$ \footnote{Another
possibility discussed in Refs.~\cite{sen3,Gab1} is
to consider the orbifold ${\rm IIB}/(-1)^{F_L}\,{\cal I}_4$
where $F_L$ is the left space-time fermion number 
and ${\cal I}_4$ is the space-time parity in four
directions.}. To see whether or not the open string
tachyon is removed by the $\Omega$ projection, we have
to analyze the interaction amplitude of the D-particle
with itself due to the exchange of unoriented
strings. To this end, we follow the general
procedure discussed in Ref.~\cite{crosscap,PCAI}, and add
to the boundary state $\ket{\widetilde B}$
the so-called 
crosscap state $\ket{C}$ which
inserts a boundary on the string world-sheet 
with opposite points being identified.
Thus, we consider
\begin{equation}
\ket{\widetilde{D0}} \equiv
\frac{1}{\sqrt{2}}\,\Big[|\widetilde{B}\rangle +
|C\rangle \Big]~~,
\label{d0I}
\end{equation}
where the factor of $1/\sqrt{2}$ has been introduced
to have the proper normalization.

Like for a usual boundary, also for a
crosscap there are two possible implementations
of the overlap equations that correspond to two 
states $\ket{C,\pm}$. However, only a linear
combination of them is selected by the GSO projection.
Since for the case at hand the R-R sector does not
play any role, we will concentrate only in the NS-NS
sector where one finds
\begin{equation}
|C\rangle = P_{\rm GSO} |C,+\rangle_{\rm NS}
=\frac{1}{2}\Big[\ket{C,+}_{\rm NS} - 
\ket{C,-}_{\rm NS}\Big]~~.
\label{crossgso}
\end{equation}
The explicit expression for the crosscap states
is \footnote{Again we only consider the matter part.
For the ghost and superghost contribution see for
example Ref.~\cite{PCAI}.}
\begin{eqnarray}
\ket{C,\pm}_{\rm NS} &=&\ii\,2^5\, {T_9 \over 2}\, 
{\rm exp} \left[ - \sum_{n=1}^{\infty} 
\frac{{\rm e}^{\ii\pi n}}{n} \alpha_{-n}\cdot S^{(9)}
\cdot\tilde{\alpha}_{-n} \right] 
\nonumber\\
&&{\rm exp} \left[ \pm\ii \sum_{r=1/2}^{\infty} 
{\rm e}^{\ii\pi r} \psi_{-r}\cdot S^{(9)}\cdot
\tilde{\psi}_{-r} 
\right] \prod_{\mu =0}^{9} | k^{\mu}=0 \rangle 
\label{crosscap}
\end{eqnarray}
where $S^{(9)}_{\alpha\beta}=\eta_{\alpha\beta}$, 
$T_9$ is given in \eq{tension} for $p=9$ and the
overall factor of $\ii$ has been introduced
for later convenience. 
Like the D9-brane, also the crosscap is a source for the non-trivial 
10-form potential of Type I theory \cite{PCAI}, and indeed 
the crosscap state $\ket{C}$ is strictly
related to the boundary state of a D9 brane with a 
pure imaginary radius; in fact one has
\begin{equation}
|C\rangle \sim \ii^{L_0+\tilde{L}_0}|D9\rangle~~.
\label{CD9}
\end{equation}
This remark will prove to be useful for technical purposes.
An important point we want to stress is that the 
normalization of the crosscap state in \eq{crosscap}
is completely fixed by the requirement
of tadpole cancellation of the Type I theory \cite{PCAI}.

We are now ready to study the interaction of a D-particle
with itself. This is due to exchange of closed strings that
propagate along a cylinder between two 
$|\widetilde{D0}\rangle$ states. 
This amplitude comprises three types of contributions. 
The first one is with two boundaries and corresponds
to a cylinder amplitude which is given by
${\cal A}= \frac{1}{2}\langle  
\widetilde{B}| D | \widetilde{B}\rangle$. Notice that
this amplitude is half of the corresponding one in the
Type IIB theory. The second type of contribution is
with one boundary and one crosscap and corresponds
to a M\"obius 
strip amplitude which is given by
${\cal M}=\frac{1}{2}\langle  
\widetilde{B}| D | C\rangle$. 
Of course, one has to consider also the conjugate expression
${\cal M}^{*}=\frac{1}{2}\langle  C| D | \widetilde{B}\rangle$
where the crosscap and the boundary have exchanged place. Finally, 
the third type of contribution is with two
crosscaps and corresponds to the Klein
bottle amplitude given by
${\cal K}=\frac{1}{2}\langle C|P|C\rangle$. This last contribution
does not contain open string poles and
refers only to the propagation of unoriented closed strings.
For these reasons we shall not consider it in our analysis.

Using the explicit expressions for the boundary and
the crosscap states, we find that the cylinder 
amplitude is
\begin{equation}
{\cal A} = {V \over 2\pi} \,
(8\pi^2 \alpha')^{-1/2}\int_{0}^{\infty} dt  
\left( {\pi \over t}\right)^{9\over 2} 
\left[  {f_3^{8}(q) \over  f_1^{8}(q)}- {f_4^{8}(q) 
\over   f_1^{8}(q)} \right]
\label{cyl1}
\end{equation}
where $q= {\rm e}^{-t}$, while the M\"obius strip 
amplitude is 
\begin{equation}
{\cal M} = 2^{9/2}\,{V \over 2\pi} \, 
(8\pi^2 \alpha')^{-1/2}\int_{0}^{\infty} dt   
\left[  {f_3^{9}(\ii\,q)\,f_1(\ii\,q) \over  
f_2^{9}(\ii\,q)\, f_4(\ii\,q)}-{f_4^{9}(\ii\,q)\,f_1(\ii\,q)  
\over   f_2^{9}(\ii\,q)\,f_3(\ii\,q)} \right]~~.
\label{moeb1}
\end{equation} 
A remark is in order here. Since the crosscap
state is related to the boundary state of a D9 brane
as we have seen in \eq{CD9}, the M\"obius 
amplitude corresponds to a system with one NN direction 
and nine DN directions which cannot be 
studied in the light cone 
gauge. Hence in this case, the use of a
covariant formalism is necessary. 

To obtain the open string channel, we must perform the modular
transformation $t\to 1/t$ in the previous expressions
so that ${\cal A}$ and ${\cal M}$ become the planar
and non-planar one-loop amplitudes of the open
strings living on the D-particle world-line.
The rules for the modular transformations of the functions $f_k$ 
with a real argument
are well known (see for example Ref.~\cite{lectPOL}). The analogous 
rules for a pure imaginary argument are instead less known, 
and thus we list them below
\begin{eqnarray}
f_1(\ii\,{\rm e}^{-{\pi s}}) &=& (2s)^{- {1/2}} 
f_1(\ii\,{\rm e}^{-{\pi\over 4s}})~~,\nonumber\\ 
f_2(\ii\,{\rm e}^{-{\pi s}}) &=& 
f_2(\ii\,{\rm e}^{-{\pi \over 4s}})~~,
\nonumber\\
f_3(\ii\,{\rm e}^{-{\pi s}}) &=& {\rm e}^{\ii {\pi/ 8}}
f_4(\ii\,{\rm e}^{- {\pi\over 4s}})~~,\nonumber\\
f_4(\ii\,{\rm e}^{-{\pi s}}) &=& {\rm e}^{-\ii {\pi/ 8}}
f_3(\ii\,{\rm e}^{- {\pi \over 4s}})~~.
\label{modular}
\end{eqnarray}
These results for $f_1$ and $f_2$ may be obtained using the rules of 
modular transformation for $f_k({\rm e}^{-\pi s})$ and elementary 
algebraic manipulations. In order to fix the phases appearing in
the transformation rules of $f_3$ and $f_4$ we have used Euler's 
pentagonal identity.

Performing the modular transformation, we then obtain 
\begin{equation}
{\cal A} = V\, 
(8\pi^2 \alpha')^{-1/2}\int_{0}^{\infty} 
{ds \over 2s}\,s^{-1/2} \left[  {f_3^{8}(r) 
\over  f_1^{8}(r)}- {f_2^{8}(r) \over   
f_1^{8}(r)} \right]
\label{annop}
\end{equation}
and
\begin{equation}
{\cal M} = 2^3 \,V\, 
(8\pi^2 \alpha')^{-1/2}\int_{0}^{\infty} 
{ds \over 2s}\,s^{-{1/2}}  
\left[{\rm e}^{-\ii\pi/4} \,{f_3^{9}(\ii\,r)f_1(\ii\,r) 
\over  f_2^{9}(\ii\,r) f_4(\ii\,r)}-{\rm e}^{\ii\pi/4}\,
{f_4^{9}(\ii\,r)f_1(\ii\,r)  \over  
 f_2^{9}(\ii\,r)f_3(\ii\,r)} \right]
\label{moebop}
\end{equation}
where $r= {\rm e}^{-\pi s}$. Notice that the crucial phases in 
\eq{moebop} are a direct consequence of the phases appearing 
in the modular transformations (\ref{modular}).
The total open string amplitude is then
\begin{equation}
{\cal A}_{open}= {\cal A}
+{\cal M}+{\cal M}^*
\label{openampl}
\end{equation}
which can be written as follows
\begin{equation}
{\cal A}_{open} =2\,V\, 
(8\pi^2 \alpha')^{-1/2}\int_{0}^{\infty} 
{ds \over 2s} \,s^{-1/2} 
\left( {\rm Z}_{{\rm NS}+}^{I}(r)+
{\rm Z}_{{\rm NS}-}^{I}(r)-
{\rm Z}_{{\rm R}}^{I}(r)\right)
\label{openampl1}
\end{equation}
where the various terms in the integrand are
\begin{eqnarray}
{\rm Z}_{{\rm NS}-}^{I}(r)
&=&  
{f_3^{8}(r)+ f_4^{8}(r) \over 4 f_1^{8}(r)} -
\ii\,2^{9 \over 2} \left(  {f_3^{9}(\ii\,r)\,f_1(\ii\,r) 
\over 4 f_2^{9}(\ii\,r)\,f_4(\ii\,r)}  +
{f_4^{9}(\ii\,r)\,f_1(\ii\,r) \over 
4 f_2^{9}(\ii\,r)\,f_3(\ii\,r)} \right) ~~,
\label{zns-}\\
{\rm Z}_{{\rm NS}+}^{I}(r)
&=&  
{f_3^{8}(r)- f_4^{8}(r) \over 4 f_1^{8}(r)} 
+2^{9 \over 2} \left(  {f_3^{9}(\ii\,r)\,f_1(\ii\,r) 
\over 4 f_2^{9}(\ii\,r)\,f_4(\ii\,r)} -
{f_4^{9}(\ii\,r)\,f_1(\ii\,r) \over 
4 f_2^{9}(\ii\,r)\,f_3(\ii\,r)} 
\right)~~,
\label{zns+}
\end{eqnarray}
and 
\begin{equation}
{\rm Z}_{{\rm R}}^{I}(r) =
{f_2^{8}(r) \over 2 f_1^{8}(r)}~~.
\label{zr}
\end{equation}
Notice that both ${\rm Z}_{{\rm NS}+}^{I}(r)$ 
and ${\rm Z}_{{\rm NS}-}^{I}(r)$ are real.
The reason to introduce these quantities is that
they can be interpreted as the partition functions
in the various sectors of the open strings
suspended between two Type I D-particles. 
To see this, we rely on the
analysis of Ref.~\cite{sen5} where the rules for 
the interactions of these open strings are
specified and shown to be consistent. Let us briefly recall
the main points of this analysis that are relevant for our
purposes.
In the NS sector, the world-sheet fermion number $F$ acts in the
usual manner both on the oscillators and on the Fock vacuum which is
taken to be odd under $(-1)^F$. The
world-sheet parity $\Omega$ has the following action
on the various string oscillators
(see for example Ref.~\cite{GIMPOL})
\begin{equation}
\alpha_{n}^{\mu} \rightarrow 
\pm {\rm e}^{\ii \pi n}\alpha_{n}^{\mu}\quad, \quad
\psi_{r}^{\mu} \rightarrow \pm {\rm e}^{\ii \pi r}
\psi_{r}^{\mu}
\label{omegaosc}
\end{equation}
where the upper (lower) sign holds for NN 
(DD) directions. The action of $\Omega$ on the 
Fock vacuum in the odd $(-1)^F$ NS sector
is taken to be 
\begin{equation}
\Omega|0\rangle^{(odd)} =-|0\rangle^{(odd)} ~~,
\label{omegavac}
\end{equation} 
while for the even $(-1)^F$ NS sector, where 
the states contain
an odd number of fermionic oscillators 
under which $\Omega$ acts as $\pm \ii$, it is taken to
be 
\begin{equation}
\Omega|0\rangle^{(even)} =-\ii\,|0\rangle^{(even)} ~~.
\label{omegavac1}
\end{equation}
According to these rules, it is not difficult to show that
the expressions in Eqs. (\ref{zns-}) and (\ref{zns+}) may
be reinterpreted as
\begin{equation}
{\rm Z}_{{\rm NS}\pm}^{I}(r) = 
{\rm Tr}_{{\rm NS}}\left( r^{(2L_0-1)}
{1\pm (-1)^{F}\over 2}{1+ \Omega 
\over 2}\right)~~,
\end{equation}
thus proving our previous statement.
Similarly, in the R sector one can show that
\begin{equation}
{\rm Z}_{{\rm R}}^{I}(r) = \frac{1}{2}\,
{\rm Tr}_{{\rm R}}\left( r^{2L_0} \right)=
{\rm Tr}_{{\rm R}}\left( r^{2L_0}{1+ 
\Omega \over 2}\right)~~,
\end{equation}
where in the last step we have used the fact that
$\Omega$ acts as a product of $\Gamma$ matrices on the
spinor R vacuum, so that ${\rm Tr}_{{\rm R}}\left( r^{2L_0}
\Omega \right)$ vanishes due to the presence of fermionic zero-modes.

Using these results, we can conclude that the
total amplitude (\ref{openampl1}) is
\begin{equation}
{\cal A}_{open} =2\,V\, 
(8\pi^2 \alpha')^{-1/2}\int_{0}^{\infty} 
{ds \over 2s} \,s^{-1/2} 
\left[{\rm Tr}_{\rm NS}\left( r^{(2L_0-1)}
{1+ \Omega \over 2}\right) - 
{\rm Tr}_{{\rm R}}\left( r^{2L_0}
{1+ \Omega \over 2}\right) \right]~~,
\label{openamplfin}
\end{equation}
which is the one-loop vacuum energy of unoriented open strings 
without any GSO projection. 
Despite this fact, there is no instability in
the system because the 
$\Omega$ projection removes 
the tachyon from the spectrum. This can be 
seen from the explicit expression of ${\rm Z}_{{\rm NS}-}^{I}(r)$ 
corresponding to the odd $(-1)^F$ NS sector
to which the tachyon belongs. 
Indeed, by expanding \eq{zns-} in powers of $r$, one
finds that
\begin{equation}
{\rm Z}_{{\rm NS}-}^{I}(r) = {\cal O}(r)
\label{expZNSm}
\end{equation}
which shows that the coefficient of the term $r^{-1}$ associated with 
the tachyon is vanishing. Notice that
for this compensation to occur, 
it is crucial that the tension of the D-particle is fixed at the 
precise value $\sqrt{2}\,T_0$. Therefore, this analysis 
provides a further check of the extra factor of 
$\sqrt{2}$ found in Ref.~\cite{sen4}. 

The massless states in the spectrum of the open
strings living on the Type I D-particle account
for its degeneracy. To count such zero-modes,
we can use the explicit
expressions for the partition
functions ${\rm Z}_{\rm NS}^I$ and ${\rm Z}_{\rm R}^I$
that we have previously derived. \eq{expZNSm}
shows that there are no massless states in the $(-1)^F$ odd
part of the NS sector. On the contrary, the
$(-1)^F$ even part of the NS sector contains nine bosonic
zero-modes, as we can see by expanding \eq{expZNSp}
in powers of $r$:
\begin{equation}
{\rm Z}_{{\rm NS}+}^{I}(r) = 9+{\cal O}(r)~~.
\label{expZNSp}
\end{equation}
These nine massless modes correspond to the freedom
of translating the D-particle along its nine
transverse directions. Finally, from \eq{zr}
we see that the R sector contains
eight non-chiral fermionic zero-modes which upon quantization
lead to a $2^8=256$ degeneracy for the
D-particle, as it should be expected for a
non-BPS multiplet of the N=1 supersymmetry
algebra in ten dimensions.

We conclude by mentioning that the D-particle
is a completely stable configuration of the Type I theory
because no tachyons develop in the spectrum of the open strings
stretching between the D-particle and the thirty-two
D9-branes which form the background.
In the boundary state formalism, this can be checked
by evaluating the amplitude
$\langle \widetilde{B}| D | D9\rangle $ and performing the modular
transformation to obtain the open
string channel. The absence of these
tachyons is also clear from the evaluation of the intercept for the
open strings in the NS sector which is
$a_{\rm NS} = {1\over 2}-{\nu\over 8}$ where $\nu$ is the
number of mixed Dirichlet-Neumann directions. In the
case at hand, $\nu=9$ and the intercept $a_{\rm NS}$ is strictly
negative so that the NS sector contains only massive modes.
As usual, there are no problems from
the R sector which contains both massless and
massive modes, the former
accounting for the SO(32) spinor
representation carried by the non-BPS D-particle.

\vskip 1.5cm
\sect{Stable D$p$-branes in the Type I theory}
\label{DpI}

In the previous sections we have shown that in the
Type IIB theory there is a non-supersymmetric D-particle
whose boundary state $\ket{\widetilde B}$ does not have any R-R part.
This unstable configuration of Type IIB becomes stable if one
performs the $\Omega$ projection to obtain the Type I theory. In this section
we want to see whether this construction can be generalized to investigate
the possibility that other non-BPS but stable D$p$-branes exist in Type I.

In order to cancel the R-R part in a boundary state the obvious
thing to do is to consider a superposition of a $p$-brane with an
anti $p$-brane (or $\bar p$-brane for short), and thus consider the boundary
state
\begin{equation}
\ket{\widetilde{Bp}} = \ket{Dp} + \ket{D{\bar p}}
\label{ppbar}
\end{equation}
which clearly has only the NS-NS component. Since we are in a Type IIB
context, $p$ has to be odd. As is well known \cite{BanksSuss}, 
this configuration is unstable due to the presence of tachyons 
in the open strings stretching between the brane and the anti brane.
To see whether or not these tachyons disappear in the 
Type I theory, we have to study their behavior under the world-sheet parity
$\Omega$. To this end, we first recall that $\Omega$ does not change
the R-R charge of a D$p$-brane if $p=1,5,9$, whereas it reverses its sign if
$p=-1,3,7$. Thus we have the following rules
\begin{equation}
\Omega(p) = p \quad,\quad \Omega(\bar p) = \bar p
\label{omegap}
\end{equation}
for $p=1,5,9$, and
\begin{equation}
\Omega(p) = \bar p \quad,\quad \Omega(\bar p) = p
\label{omegap1}
\end{equation}
for $p=-1,3,7$.
The tachyons we are concerned about, are $(-1)^F$ odd states in the
$p$-$\bar p$ and $\bar p$-$p$ sectors of the open strings
living on the brane world-volume. The
vacuum states out of which they are constructed are therefore
\begin{equation}
\ket{0}_{p\bar p} = \ket{0}\,\otimes\,
\Lambda_{p\bar p}
\quad ~\mbox{and}~ \quad
\ket{0}_{\bar p p} = \ket{0}\,\otimes\,
\Lambda_{\bar p p}
\label{vacuapp}
\end{equation}
where $\ket{0}$ is the NS Fock vacuum in the $-1$ superghost picture,
and $\Lambda_{p\bar p}$ and $\Lambda_{\bar p p}$ are
the Chan-Paton factors which label all states of
the $p$-$\bar p$ and $\bar p$-$p$ sectors respectively.
The operator $\Omega$ acts on the Fock vacuum in the
usual manner, {\it i.e.} $\Omega\,\ket{0}=-\ii\,\ket{0}$.
The action of $\Omega$ on the Chan-Paton factors
is instead more subtle and is given by
\begin{equation}
\Omega\,\Lambda_{p\bar p}\,\Omega^{-1} = \omega_p\,
\Lambda_{\Omega(\bar p)\Omega(p)}
\label{omegalambda}
\end{equation}
where $\omega_p$ is a suitable phase.
From \eq{omegap}, we see that for $p=1,5,9,$ 
$\Omega$ maps $\Lambda_{p \bar p}$ into $\Lambda_{\bar p p}$
and vice versa. Since $\Omega$ 
relates the $p$-$\bar p$ and the $\bar p$-$p$ sectors, it cannot 
be used to remove the tachyons, and thus we conclude that for $p=1,5,9,$ 
the $p$-$\bar p$ system described by the boundary state
(\ref{ppbar}) does
not yield any stable configuration.

Things are different if $p=-1,3,7$. In this case, \eq{omegap1}
shows that $\Omega$ does not exchange
$\Lambda_{p\bar p}$ and $\Lambda_{\bar p p}$; furthermore
using the arguments
of Refs.~\cite{GIMPOL}, one can prove that
the intrisic phase of \eq{omegalambda} is
$\omega_p=(-\ii)^{\frac{9-p}{2}}$,
so that the states $\ket{0}_{p\bar p}$ and
$\ket{0}_{\bar p p}$ are eigenstates of $\Omega$ with eigenvalues
$(-\ii)^{\frac{11-p}{2}}$. Since $\Omega^2=1$, it
is conceivable that the $\Omega$ projection can eliminate the tachyons
and stabilize the $p$-$\bar p$ systems when $p=-1,3,7$ \cite{WITTEN}.

As we have seen in \secn{d1antid1}, there is a less obvious
way to remove the R-R part of a boundary state, namely
by considering the superposition of a brane and an anti-brane
in the presence of a ${\bf Z}_2$ Wilson line \cite{sen4}.
The boundary state corresponding to this configuration is
\begin{equation}
\ket{Dp}+\ket{D\bar p\,'}
\label{sup'}
\end{equation}
where $p$ is odd, and the $'$ denotes the Wilson line on the anti brane.
As we have seen, the R-R part of this superposition does not
trivially vanish, but after tachyon condensation it does. The resulting
configuration is an unstable D$(p-1)$-brane of the Type IIB
theory whose boundary state
is related to the combination \eq{sup'} by the discrete transformation
${\cal T}$ described in \secn{d1antid1}
and contains only a NS-NS component. The instability of these
even D-branes is due to
the presence of tachyons in the spectrum of the open strings
living in their world-volume. For simplicity, we refer to
these strings as
the $p$-$p$ strings (where now $p$ is even!).
The tachyons are then in the $(-1)^F$ odd part
of the spectrum of the $p$-$p$ strings, and are constructed from
a vacuum which we denote by $\ket{0}_{pp}^{(odd)}$.
To decide about the fate of such tachyons
in the Type I theory, we have
to define the action of $\Omega$ on the
spectrum of the $p$-$p$
strings. $\Omega$ acts as usual on the oscillators
(see \eq{omegaosc}), whereas, using the arguments of
Ref.~\cite{GIMPOL} one can show that
\begin{equation}
\Omega \,\ket{0}_{p p}^{(odd)} =
-(\ii)^{ p \over 2} \,\ket{0}_{p p}^{(odd)}~~. 
\label{evenptac}
\end{equation}
From this equation, we see that for $p=2,6,$ $\Omega^2=-1$
on the vacuum, and hence the operator $(1+\Omega)/2$
cannot be used to truncate the spectrum and remove
the tachyons. On the contrary, there are no obstructions in performing
the $\Omega$ projection in the other cases, {\it i.e.} $p=0,4,8.$

This analysis leads us to consider only the 
Type IIB non-BPS D$p$-branes with 
$p=-1,0,3,4,7,$ $8$, since these are the only ones that
in principle can become stable in Type I.
Correspondingly, for these values of $p$ we
consider GSO projected boundary 
states $\ket{\widetilde{Bp}}$ that contain only a NS-NS
part and are given by
\begin{equation}
\ket{\widetilde{Bp}} = \frac{1}{2} 
\left( \ket{\widetilde{Bp}, +} - 
\ket{\widetilde{Bp}, -} \right)
\label{pgso2}
\end{equation}
where 
\begin{equation}
\ket{\widetilde{Bp}, \pm} = 
{\mu_p T_p \over 2}\,
\ket{{Dp}_{\rm mat}, \pm}_{\rm NS}
\label{tens}
\end{equation}
in the notation of \secn{boundarystate}. 
In this last equation we have 
introduced a {\it positive} parameter $\mu_p$ to renormalize the
brane tension, which will be fixed later by imposing the
stability condition, {\it i.e.} by requiring the 
cancellation of tachyons in the Type I theory.

We now proceed as in \secn{typeI}: we first add to the boundary
state $\ket{\widetilde{Bp}}$ the crosscap state $\ket{C}$
\begin{equation}
\ket{\widetilde{Dp}}= \frac{1}{\sqrt{2}}
\Big[\ket{\widetilde{Bp}} + \ket{C}\Big]~~,
\label{ptypeI}
\end{equation}
and then study the interaction amplitude corresponding
to the exchange of closed strings between
two $\ket{\widetilde{Dp}}$ states. For our purposes,
we consider only the contributions with open string poles,
namely the cylinder amplitude and the M\"obius strip amplitude
given by
\begin{equation}
{\cal A} = \frac{1}{2}\,\bra{\widetilde{Bp}} D \ket{\widetilde{Bp}} \quad \mbox{and}
\quad
{\cal M} = \frac{1}{2}\,\bra{\widetilde{Bp}} D \ket{C}~~.
\label{amplo}
\end{equation}
Using the explicit expressions of the boundary and crosscap
states, after standard manipulations, we find
\begin{equation}
{\cal A} = \frac{\mu_p^2}{2}\, 
{V_{p+1} \over 2 \pi} \,(8 \pi^2 \a')^{- {p+1 \over 
2}} \int_0^{\infty}  dt \, \left( {\pi \over t} \right)^{\nu \over 2}
\left[ { f_3^8(q) - f_4^8 (q) \over f_1^8(q)} \right]
\label{annp}
\end{equation}
and
\begin{equation}
{\cal M} =
2^4 \, \mu_p  {V_{p+1} \over 2 \pi} \,
(8 \pi^2 \a')^{- {p+1 \over 2}} \int_0^{\infty} dt 
\left[ {f_4^{8-\nu}(\ii\,q) \,f_3^{\nu}(\ii\,q) 
\over f_1^{8 - \nu}(\ii\,q)\,f_2^{\nu}(\ii\,q) } -
{f_3^{8- \nu} (\ii\,q)\, f_4^{\nu}(\ii\,q) \over f_1^{8- \nu}
(\ii\,q)\,f_2^{\nu}(\ii\,q)} \right]
\label{moebp}
\end{equation}
where $q={\rm e}^{-t}$ and $\nu=9-p$. Note that if
$p=0$ and $\mu_0=\sqrt{2}$ these expressions reduce
to the amplitudes (\ref{cyl1}) and (\ref{moeb1}).

To obtain the open string channel we perform
the modular transformations 
$t\to\pi/ s$ in ${\cal A}$ and $t\to \pi/4 s$ in ${\cal M}$, and
after using the modular properties of the functions $f_k$ 
(see
{\it e.g.} \eq{modular}), we find that Eqs. (\ref{annp}) and
(\ref{moebp}) become respectively
\begin{equation}
{\cal A}= \frac{\mu_p^2}{2}\, V_{p+1} \,(8 \pi^2 \a')^{- {p+1 \over 2}} 
\int_0^{\infty}  {ds \over 2s} \, s^{- {p+1 \over 2}} 
\left[ { f_3^8(r) - f_2^8 (r) \over f_1^8(r)} \right] 
\label{anulus}
\end{equation}
and 
\begin{eqnarray}
{\cal M} &=&  2^{5-p \over 2} \, \mu_p \, V_{p+1} \,
 (8 \pi^2 \a')^{- {p+1 \over 2}}
\int_0^{\infty} {ds \over 2s} \, s^{- {p+1 \over 2}}
\left[ {\rm e}^{-\ii \nu \pi/4}\, {f_4^{8- \nu}(\ii\,r)\,f_3^{\nu}(\ii\,r)
 \over f_1^{8 - \nu}(\ii\,r)\,f_2^{\nu}(\ii\,r) } \right. 
 \nonumber \\
&&~~~~~~~~~~~~~~~~~
\left.- {\rm e}^{\ii \nu \pi/4}\,{f_3^{8- \nu} (\ii\,r)\, f_4^{\nu}(\ii\,r)
 \over f_1^{8- \nu} (\ii\,r)\,f_2^{\nu}(\ii\,r)} \right]
\label{moebius}
\end{eqnarray}
where $r = {\rm e}^{- \pi s}$.

The spectrum of the open strings living on the world-volume
of these Type I D-branes can be analyzed by expanding the total
amplitude ${\cal A}_{open}={\cal A}+{\cal M}+{\cal M}^\ast$
in powers of $r$. The leading term in this expansion is
\begin{equation}
{\cal A}_{open} \sim \int_0^{\infty} {ds \over 2s}\, 
s^{- {p+1 \over 2}}
\left[ \mu_p^2 - 2 \mu_p \sin ( {\pi \over 4} \nu) \right] 
r^{-1}~~.
\label{asymptotic}
\end{equation}
The $r^{-1}$ behavior of the integrand signals the presence
of tachyons in the spectrum; 
therefore, in order not to have them, we must 
require that 
\begin{equation}
\mu_p = 2\,\sin (\frac{\pi}{4}\nu)~~.
\label{stability}
\end{equation}
Since $\nu=9-p$ and $\mu_p$ has to be positive,
there is no solution if $p=3,4$; in the other cases
we find
\begin{equation}
 \begin{array}{|c|c|}\hline
 p         & \mu_p       \\ \hline
 -1       & 2             \\ \hline
  0        & \sqrt{2}    \\ \hline
    7        & 2               \\ \hline
  8        & \sqrt{2}     \\ \hline
 \end{array}
\label{tabular}
\end{equation}
From this table we see that there exist two
even non-BPS but stable D$p$-branes: 
the D-particle and the D8-brane \cite{WITTEN}. 
Both of them have a tension that is a factor
of $\sqrt{2}$ bigger than the corresponding branes of the
Type IIA theory. This is in agreement
with the result of the explicit construction presented in
Sections 3 and 4.
Moreover, there exist two 
odd non-BPS but stable D$p$-branes: the D-instanton and the D7-brane
\cite{WITTEN}.
Their tension is twice the one of the corresponding
Type IIB branes, in accordance with the fact that they
can be simply interpreted as the superposition of a brane
with an anti brane, so that the R-R part of the boundary state
cancels while the NS-NS part doubles.

It is interesting to see what happens in this
construction if one performs the $\Omega$
projection by using a symplectic crosscap state
instead of the orthogonal one considered so far.
This would correspond to quantize the thirty-two
D9-branes of the background
with symplectic rather than orthogonal
Chan-Paton factors, a procedure which is known
not to be fully consistent but which can nevertheless 
be considered up to some extent \cite{WITTEN}.
In this case everything proceeds as before, except
that the M\"obius strip contributions have a different
sign. Thus the tachyon cancellation condition (\ref{stability})
gets replaced by
\begin{equation}
\mu_p = -2\,\sin (\frac{\pi}{4}\nu)~~.
\label{stability1}
\end{equation}
Since $\mu_p$ has to be positive, this equation 
has solution only if $p=3,4$, 
in which case we get
\begin{equation}
 \begin{array}{|c|c|}\hline
 p         & \mu_p       \\ \hline
   {}{}3        & 2            \\ \hline
  {}{}4        & \sqrt{2}  \\ \hline
   \end{array}
\label{tabular1}
\end{equation}
The classification of the stable non-BPS D-branes of Type I based
on tables (\ref{tabular}) and (\ref{tabular1}) is in
complete agreement with the results of Refs.~\cite{WITTEN,HORAVAK,GUKK}
derived from the K-theory of space-time.

We end by observing that actually only the
D-instanton and the D-particle are fully stable
configurations of the Type I theory. In fact, in all
other cases there are tachyons that develop in the
spectrum of the open strings ending on the thirty-two
D9-branes that form the Type I background.
It would be very interesting to study and analyze in more
detail the properties and the interactions of these non-BPS but
stable D-branes. We hope that the boundary state formalism 
discussed in this paper might be useful for this purpose.

\vskip 2cm
{\large {\bf {Acknowledgements}}}
\vskip 0.5cm
We would like to thank P. Di Vecchia, I. Pesando and R. Russo
for many useful discussions.

\vskip 2cm
\appendix{\Large {\bf {Appendix A}}}
\label{appa}
\vskip 0.5cm
\renewcommand{\theequation}{A.\arabic{equation}}
\setcounter{equation}{0}
\noindent
In this appendix we compute the partition function 
of the open strings living on the world-wolume 
of the D-string -- anti D-string pair defined in
\secn{d1antid1} at the critical radius $R_c=\sqrt{\a '/2}$
and for a generic value $\theta$ of the 
tachyon v.e.v. 
We present our calculation in the Type IIB theory. 
This open string partition function can be obtained by performing
a modular transformation $t\to 1/t$  on the amplitudes 
(\ref{vacampl}) and (\ref{vacampl1}) that we computed using
the boundary state formalism. 
As usual, in going from the open to the closed string
channel we obtain the following identification of the various sectors: 
the NS sector of the open string 
comes from the NS-NS sector of the closed string, 
the NS$(-)^F$ from the R-R and the R from the 
NS-NS$(-)^F$. Performing the modular transformation
on the sum of Eqs. (\ref{vacampl}) and (\ref{vacampl1}), after
standard manipulations we get
\begin{equation}
{\cal A}(\theta) = 2\,V\, 
(8\pi^2 \alpha')^{-1/2}\int_{0}^{\infty} 
{ds \over 2s} \,s^{-1/2} 
\Big( {\rm Z}_{{\rm NS}}(r;\theta) -
{\rm Z}_{{\rm R}}(r;\theta)\Big)
\label{appa1}
\end{equation}
where
\begin{eqnarray}
{\rm Z}_{{\rm NS}}(r;\theta ) &=& 
\sum_{n} 
r^{4n^2} \,{f_3^{8}(r)-f_4^{8}(r) \over 2 f_1^{8}(r)} + 
\sum_{n} r^{4(n+{1 \over 2})^2} \,
{f_3^{8}(r)+f_4^{8}(r) \over 2 f_1^{8}(r)}\nonumber\\
&+&\sum_{n} r^{4(n+{\theta \over 2})^2} \,
{f_3^{8}(r)-f_4^{8}(r) \over 2 f_1^{8}(r)}
+\sum_{n} r^{4(n+{1+\theta  \over 2})^2} \,
{f_3^{8}(r)+f_4^{8}(r) \over 2 f_1^{8}(r)}
\label{appa2}
\end{eqnarray}
and 
\begin{equation}
{\rm Z}_{{\rm R}}(r;\theta ) = \sqrt{2} \,{f_2^{7}(r)\,f_3(r) 
\over  f_1^{7}(r)\,f_4(r)}
\label{appa3}
\end{equation}
with $r={\rm e}^{-\pi s}$. Notice that ${\rm Z}_{{\rm R}}(r;\theta )$
is actually independent of $\theta$.
Eqs. (\ref{appa2}) and (\ref{appa3}) are precisely
the open string partition functions in the NS and R sectors
that can be evaluated from the rules given 
in Ref.~\cite{sen4}. 
In particular, the R sector partition
function (\ref{appa3}) is
obtained using as degrees of freedom for the compact
direction the fields $(\phi , \eta )$ defined in \secn{d1antid1}.
In this representation, the contribution of the
compactified direction is that of a
boson with mixed Neumann-Dirichlet 
boundary conditions (yielding the $f_4$ in the denominator
of \eq{appa3}) and of a NS fermion (yielding the $f_3$ in the
numerator). Being independent of $\theta$
this partition function can also be computed at $\theta=0$ using 
the original $(X,\psi )$ degrees of
freedom. In this representation
we have the contribution of a compactified boson 
and of a Ramond fermion so that
\begin{equation}
{\rm Z}_{{\rm R}}(r;\theta ) = 
\sum_{n} r^{n^2} \,{f_2^{8}(r) \over f_1^{8}(r)}~~.
\label{appa4}
\end{equation}
The equality between Eqs. (\ref{appa3}) and (\ref{appa4})
is ensured by the identities (\ref{ident}). 

If we set $\theta =1$, we observe a remarkable  
simplification in the NS part of the partition 
function which becomes
\begin{equation}
{\rm Z}_{{\rm NS}}(r;\theta =1 ) = 
\sum_{n} r^{n^2} \,{f_3^{8}(r) \over f_1^{8}(r)}~~.
\label{appa5}
\end{equation}
Eqs. (\ref{appa5}) and (\ref{appa4}) are the partition functions 
of an open string in the NS and R sectors without any GSO projection.
\newpage
\appendix{\Large {\bf {Appendix B}}}
\label{appb}
\vskip 0.5cm
\renewcommand{\theequation}{B.\arabic{equation}}
\setcounter{equation}{0}
\noindent
In this appendix we show that the boundary state
$\ket{\widetilde B,+}$ given in \eq{newbound} is
equivalent to the boundary state $\ket{B(\theta),+}_{\rm NS}$
given in \eq{hatbns} for $\theta=1$.  To prove this 
equivalence we consider a few closed string states 
and check that
they have the same overlap with both $\ket{\widetilde B,+}$ 
and $\ket{B(\theta=1),+}_{\rm NS}$. Since the 
difference between these two boundary states 
is only in the compact direction, in the 
following we will focus only on its contribution 
and understand all the rest.

We begin with the following state
\begin{equation}
\ket{\chi} = \ket{n=1,w=0} + \ket{n=-1,w=0}
\label{state1}
\end{equation}
which is created by the vertex operator
\begin{equation}
W_{\chi}(z,\bar z) = {\rm e}^{\frac{\ii}{\sqrt{2\a'}}\left[
X_L(z) + X_R(\bar z)\right]} +
{\rm e}^{-\frac{\ii}{\sqrt{2\a'}}\left[X_L(z)+X_R(\bar z)\right]}~~.
\label{vertex1}
\end{equation}
When we saturate the boundary state $\ket{\widetilde B,+}$
with $\ket{\chi}$, we obtain
\begin{equation}
\bra{\chi}\widetilde B,+\rangle = 2\,{\cal N}\,\Phi
\label{result1}
\end{equation}
where we have used \eq{norm} and absorbed in ${\cal N}$
all factors coming from the normalization of the boundary state
and from the contributions of non-compact directions.

To compute the overlap between $\ket{\chi}$ 
and $\ket{B(\theta),+}_{\rm NS}$, we have first to find 
the representation of $\ket{\chi}$ in terms of the fields
$(\phi,\eta)$. To this end, we use the bosonization formulas
of \secn{d1antid1} to write the vertex operator
$W_{\chi}$ as follows
\begin{eqnarray}
W_{\chi}(z,\bar z) &=& \ii\, \eta(z)\,\tilde \eta(\bar z)
- \frac{1}{2}\left\{{\rm e}^{\frac{\ii}{\sqrt{2\a'}}\left[
\phi_L(z) + \phi_R(\bar z)\right]}
+{\rm e}^{\frac{\ii}{\sqrt{2\a'}}\left[
\phi_L(z) - \phi_R(\bar z)\right]}
\right.\nonumber \\
&&\left.+
{\rm e}^{-\frac{\ii}{\sqrt{2\a'}}\left[
\phi_L(z) - \phi_R(\bar z)\right]}
+{\rm e}^{-\frac{\ii}{\sqrt{2\a'}}\left[
\phi_L(z) + \phi_R(\bar z)\right]}
\right\}~~.
\label{vertex12}
\end{eqnarray}
From this expression we see that
the $(\phi,\eta)$-representation of $\ket{\chi}$ is
\begin{eqnarray}
\ket{\chi}&=& 
\ii\,\eta_{-1/2}\,\tilde\eta_{-1/2}\ket{n_\phi=0,w_\phi=0}
-\frac{1}{2}\Big[\ket{n_\phi=1,w_\phi=0}+\ket{n_\phi=0,w_\phi=2}
\nonumber \\
&&+\ket{n_\phi=0,w_\phi=-2}+\ket{n_\phi=-1,w_\phi=0}\Big]~~.
\label{state12}
\end{eqnarray}
When we saturate $\ket{B(\theta),+}_{\rm NS}$
with this state, we get
\begin{equation}
\langle \chi\ket{B(\theta),+}_{\rm NS} = 
\left(1-\cos(\pi\theta)\right)\,{\cal N}\,\Phi
\label{result12}
\end{equation}
which agrees with \eq{result1} if $\theta=1$. Notice that for 
$\theta=0$, this overlap vanishes; this is to be expected
since $\ket{B(\theta=0),+}_{\rm NS}$
is equivalent to the original boundary
state $\ket{B,+}_{\rm NS}$ of \eq{d1ns} with which
$\ket{\chi}$ has zero overlap.

Let us now consider the following state
\begin{equation}
\ket{\rho} = \alpha_{-1}\,\tilde\alpha_{-1}\ket{n=0,w=0}
\label{state2}
\end{equation}
which corresponds to the vertex operator
\begin{equation}
W_{\rho}(z,\bar z) = -\frac{1}{2\a '}\,\partial X_L(z)\,
\bar\partial X_R(\bar z)~~.
\label{vertex2}
\end{equation}
Saturating $\ket{\rho}$ against the boundary 
state $\ket{\widetilde B,+}$ we get
\begin{equation}
\langle \rho\ket{\widetilde B,+} = {\cal N}\,\Phi~~.
\label{result2}
\end{equation}
After bosonization the vertex $W_{\rho}$ becomes
\begin{eqnarray}
W_{\rho}(z,\bar z) &=& -\frac{\ii}{2}\,\eta(z)
\tilde \eta(\bar z)
\left\{{\rm e}^{\frac{\ii}{\sqrt{2\a'}}\left[
\phi_L(z) + \phi_R(\bar z)\right]}
+{\rm e}^{\frac{\ii}{\sqrt{2\a'}}\left[
\phi_L(z) - \phi_R(\bar z)\right]}
\right.\nonumber \\
&&\left.+
{\rm e}^{-\frac{\ii}{\sqrt{2\a'}}\left[
\phi_L(z) - \phi_R(\bar z)\right]}
+{\rm e}^{-\frac{\ii}{\sqrt{2\a'}}\left[
\phi_L(z) + \phi_R(\bar z)\right]}
\right\}~~,
\label{vertex22}
\end{eqnarray}
and the corresponding state is
\begin{eqnarray}
\ket{\rho} &=& -\frac{\ii}{2}\,
\eta_{-1/2}\,\tilde\eta_{-1/2}\,
\Big[\ket{n_\phi=1,w_\phi=0}+\ket{n_\phi=0,w_\phi=2}
\nonumber \\
&&+\ket{n_\phi=0,w_\phi=-2}+\ket{n_\phi=-1,w_\phi=0}\Big]~~.
\label{state22}
\end{eqnarray}
Computing its overlap with the boundary state 
$\ket{B(\theta),+}_{\rm NS}$, we get
\begin{equation}
\langle \rho\ket{B(\theta),+}_{\rm NS} = - \cos(\pi\theta) \,
{\cal N}\,\Phi
\label{result22}
\end{equation}
which again agrees with \eq{result2} for $\theta=1$.

Our last example concerns the state
\begin{equation}
\ket{\sigma} = \ii\,\psi_{-1/2}\,\psi_{-1/2}
\ket{n=0,w=0}
\label{state3}
\end{equation}
which is created by the following vertex operator
\begin{equation}
W_{\sigma}(z,\bar z) = \ii\,\psi(z)\,\tilde\psi(\bar z)~~.
\label{vertex3}
\end{equation}
Its overlap with the boundary state $\ket{\widetilde B,+}$ is
\begin{equation}
\langle \sigma \ket{\widetilde B,+} = -{\cal N}\,\Phi
\label{result3}
\end{equation}
Using the bosonization rules of \secn{d1antid1}, 
one easily sees that the
vertex $W_\sigma$ written in terms of $\phi$ and $\eta$ is
\begin{eqnarray}
W_{\sigma}(z,\bar z) &=&
-\frac{1}{2}
\left\{{\rm e}^{\frac{\ii}{\sqrt{2\a'}}\left[
\phi_L(z) + \phi_R(\bar z)\right]}
-{\rm e}^{\frac{\ii}{\sqrt{2\a'}}\left[
\phi_L(z) - \phi_R(\bar z)\right]}
\right.\nonumber \\
&&\left.-
{\rm e}^{-\frac{\ii}{\sqrt{2\a'}}\left[
\phi_L(z) - \phi_R(\bar z)\right]}
+{\rm e}^{-\frac{\ii}{\sqrt{2\a'}}\left[
\phi_L(z) + \phi_R(\bar z)\right]}
\right\}~~,
\label{vertex32}
\end{eqnarray}
so that the corresponding state is
\begin{eqnarray}
\ket{\sigma} &=& -\frac{1}{2}\,
\Big[\ket{n_\phi=1,w_\phi=0}-\ket{n_\phi=0,w_\phi=2}
\nonumber \\
&&-\ket{n_\phi=0,w_\phi=-2}+\ket{n_\phi=-1,w_\phi=0}\Big]~~.
\label{state32}
\end{eqnarray}
When we compute its overlap with the boundary state
$\ket{B(\theta),+}_{\rm NS}$ we get
\begin{equation}
\langle \sigma \ket{B(\theta),+}_{\rm NS} = \cos(\pi\theta)\,{\cal N}\,\Phi
\label{result32}
\end{equation}
which once more coincides with \eq{result3} if $\theta=1$.

All these examples provide convincing evidence that indeed
the boundary state $\ket{\widetilde B,+}$ defined in \eq{newbound}
is equivalent to the boundary state $\ket{B(\theta),+}_{\rm NS}$ defined 
in \eq{hatbns} for $\theta=1$. Furthermore, as a byproduct of these 
calculations, we can also verify that $\ket{B(\theta=0),+}_{\rm NS}$ 
and the original boundary state $\ket{B,+}_{\rm NS}$ of \eq{d1ns}
are equivalent as they should.

\end{document}